\documentclass[Letter, 10pt, conference]{ieeeconf} 

\IEEEoverridecommandlockouts  

\usepackage{enumerate}
\usepackage{amssymb}
\usepackage{amsmath}
\usepackage[latin1]{inputenc}
\usepackage{mathrsfs}
\usepackage{psfrag}
\usepackage{graphics}
\usepackage{amsfonts} 
\usepackage[aboveskip=0pt,font=footnotesize]{caption}
\usepackage[usenames,dvipsnames,svgnames,table]{xcolor}
\usepackage{array}
\usepackage{tikz}
\usepackage{booktabs}
\usepackage{multirow}
\newcommand{\head}[1]{\textnormal{\textbf{#1}}}
\newcommand{\normal}[1]{\multicolumn{1}{l}{#1}}

\usepackage{epsfig} 
\usepackage{times} 
\usepackage{psfrag}

\input{psfig.tex} 

\newtheorem{theorem}{Theorem}

\newtheorem{definition}{Definition}

\newtheorem{remark}{Remark}

\def\salt{\vskip 0.5 true cm}

\input amssym.def
\input amssym

\def\Re{{\mathop{\Bbb R}}}

\def\blacksquare{\hbox{\vrule width 4pt height 4pt depth 0pt}}

\def\qedp{\hfill{$\blacksquare$}}

\def\1D#1#2{{{\partial}\over{\partial #2}}#1}

\def\1d#1#2{{{\rm d}\over{{\rm d} #2}}#1}

\def\salt{\vskip 0.1 true cm}

\def\Re{\mathop{\Bbb R}}

\def\blacksquare{\hbox{\vrule width 4pt height 4pt depth 0pt}}

\def\qedp{\hfill $\blacksquare$}

\def\1D#1#2{{{\partial}\over{\partial #2}}#1}

\def\1d#1#2{{{\rm d}\over{{\rm d} #2}}#1}

\title{Classification for Dynamical Systems: \\ Model-based Approach and Support Vector Machines 
}

\author{Giorgio Battistelli and Pietro Tesi 
\thanks{The authors are with 
        DINFO, University of Florence, 50139 Florence, Italy 
        {\tt\small \{giorgio.battistelli,pietro.tesi\}@unifi.it}.
       }
}

\begin{document}

\maketitle

\begin{abstract}
We consider the problem of classifying trajectories generated by dynamical 
systems. We investigate a model-based approach, the 
common approach in control engineering, and a data-driven approach based 
on Support Vector Machines, a popular method in the area of machine learning.
The analysis points out connections between the two approaches and 
their relative merits. 
\end{abstract}


\section{Introduction}

Assume that we are given two dynamical systems, whose underlying 
dynamics might be unknown. We are interested in designing a \emph{classifier},
that is a machine which, given an observed trajectory generated by either the systems,
correctly identifies which of the candidate systems has generated such a trajectory.
We will refer to this problem as the problem of \emph{classification for dynamical systems}.

Problems of this type are ubiquitous within the community of \emph{systems and control}.
For instance, in fault detection one system may represent the 
behavior in nominal conditions and another a prescribed 
faulty behavior. Similarly, in networked control one system
may represent the closed-loop behavior when data are
successfully transmitted, while another system may represent the 
open-loop behavior in the presence of packet dropouts.
Problems of this type naturally arise also when dealing with systems
that naturally exhibit multiple operating modes, for example switched circuits.
On the other hand, in \emph{computer science}, classification is the core of \emph{machine learning} 
for pattern recognition \cite{Vapnik1995,Burges1998}, which has found several applications 
in different fields of engineering, including applications that are 
intensively studied in control engineering like fault detection and diagnosis \cite{Mahadevan2009}.
Yet, with regard to this problem, the interaction between these two communities has been low.

Within the {systems and control} community, classification 
for dynamical systems has been studied in connection 
with the analysis of switched\,/\,multi-mode systems
\cite{ViChSoSa03}\nocite{BaEg04,VuLiberzon2008,Lou2011,Ba13}-\cite{battSCL},
often under the term \emph{mode-identification},
which is defined as the problem of reconstructing the active mode of a switched
system from its output trajectories.
The common approach is a \emph{model-based} approach:
assuming a correct model of the system for each operating mode, 
one can check whether or not mode-identification is feasible 
via dynamic-dependent conditions, and mode-identifiers (in fact, classifiers)
can be obtained in terms of rank tests, least-square functions or dynamical systems.
While the theory also generalizes to noisy observations \cite{battSCL,TanLibCDC2011}
and to some types of nonlinear dynamics \cite{TanLibAUTOMATICA2010,battTAC},
little is known on how to approach model-based classification
if one departs from the hypothesis that the dynamics of the system are known with perfect accuracy. 
Even in the simplest case where the dynamics are associated with \emph{parametric} 
uncertainty, building a classifier is a non-trivial task. The difficulty is similar to the one
encountered in adaptive control based on multiple models, where a main issue
is indeed to guarantee that modelling inaccuracies do not destroy the
learning capability of the control scheme \cite{liberzon}.

In computer science, the main paradigm to classification 
is instead the \emph{data-driven} paradigm.
Classifiers are designed by choosing a function with adjustable parameters
selected using a number of training data, called the \emph{examples}. The resulting function 
(the classifier) is then evaluated according to its capability to \emph{generalize} from the training dataset, 
that is to correctly map new examples. Popular methods are \emph{Neural Networks} (NN) 
and \emph{Support Vector Machines} (SVM), whose 
capability to {generalize} from a training dataset can be quantified 
via suitable loss functions such as the \emph{risk} function \cite{Vapnik1995}.
{Data-driven} methods have the intrinsic potential to overcome issues related to model uncertainty, 
and have already proven their effectiveness in challenging applications such as the prediction of epileptic 
seizures from recording of EEG signals \cite{Chisci2010}.
However, it is not obvious how to tailor the analysis and design
of data-driven methods to the specific context where data come from dynamical systems.
It is worth pointing out that classification of data generated 
from dynamical systems is not new in computer science. In fact, it can be
regarded as classification of \emph{time series} once we assume the existence 
of an underlying data-generating system. Yet, approaches which take
this standpoint still try to incorporate models into the learning task, either
to extract from data informative \emph{features} \cite{Brodersen2011}
or to construct suitable \emph{kernel} functions \cite{Jebara2004,Vishwanathan2007}.
While incorporating models is a natural step to take, it
leads to the previous question of how to handle model uncertainty,
and does not help to understand the performance 
achievable by model-free schemes. 
Similar issues related to model-based approaches have been pointed out 
also in the context of \emph{clustering} \cite{Lauwers2017}.

In this paper, we consider autonomous linear systems and approach the 
classification problem from both model-based and data-driven perspectives,
pointing out relative merits and establishing connections between the two.
We first consider a model-based approach and derive a classifier assuming the 
knowledge of the system dynamics. This approach has two fundamental merits: 
i) to highlight necessary conditions for the existence of a correct classifier (problem feasibility);
ii) to guide the design and analysis of a data-driven solution. 
In connection with ii), the model-based approach shows that 
under problem feasibility one can design a correct classifier 
which can be interpreted in terms of \emph{polynomial kernels} \cite{Burges1998}.
Building on this result, we consider a data-driven approach based on SVM.
By using properties stemming from the model-based solution, 
we provide bounds on the margin of the classifier 
and quantify its generalization performance \cite{Bartlett1999}
as a function of the systems one wishes to classify.  

The rest of this paper is as follows. Sections II and III formalize
the problem of interest, and recall basic concepts regarding SVM.  
In Sections IV and V, we present the main results. Section VI discusses the results and open problems. 
Numerical simulations are reported in 
Section VII, while Section VIII provides concluding remarks.
 

\section{Framework}\label{overview}

Consider two linear dynamical systems
\begin{eqnarray} \label{eq:sys}
\def\arraystretch{1.3}
\Sigma_i \, \sim \, \left\{
\begin{array}{l}
x_{i}(t+1) = A_{i} \, x_i(t)  \\ 
y_i(t)   = C_{i} \, x_i(t)
\end{array}, \quad i =1,2
\right.
\end{eqnarray}
where $t \in \mathbb Z_+ := \{0,1,\ldots\}$ denotes time; $x_i \in {\Re}^{n_i}$ 
is the state; $y_i \in {\Re}^m$ is the output; $A_i$ and $C_i$ are state and output
transition matrices. We will assume that each $\Sigma_i$
is observable (in a control-theoretic sense). 
This entails no loss of generality in that if $\Sigma_i$
is not observable, all subsequent developments apply to the 
observable subsystem obtained via a Kalman observability
decomposition.
Assume now that we are given a sequence
\begin{eqnarray} \label{eq:observ_vector}
Y \, :=\, \mbox{col} (y(0),y(1),\ldots,y(N-1))
\end{eqnarray}
of $N$ measurements generated by one of the two systems, that is
$Y = \mbox{col} (y_i(0),y_i(1),\ldots,y_i(N-1))$ with $i \in \{1,2\}$,
but we have no direct information on which of the two systems has generated $Y$.
We are interested in determining which of the two systems
has generated $Y$, referring to this problem as 
the problem of \emph{classification}.

To make the problem definition precise, let
\begin{eqnarray} \label{eq:obs_matrix}
\mathcal O_i  \, := \,
\left[
\begin{array}{c}
\, C_i \,\\
\,C_i\,A_i \, \\
\vdots \\
\,\,C_i\,(A_i)^{N-1} \,\,
\end{array}
\right ]
\end{eqnarray}
be the observability matrix of order $N$ of the pair $(C_i,A_i)$, 
and let 
\begin{eqnarray} \label{eq:set_L}
\mathcal L_i \, := \,
\left\{ \mathcal O_i x, \, x \in \mathbb R^{n_i}: \,\, \mathcal O_i x \ne 0 \right\} 
\end{eqnarray}
be the set of all possible nonzero trajectories 
of $N$ samples that can be generated by the $i$-th system.
\salt

\begin{definition}[Classifiers and correctness] \label{def:class}
A \emph{classifier} is any function $f: \mathcal Y \rightarrow \mathbb R$, where 
$\mathcal Y$ is the space of the input data. 
A {classifier} for the dynamical systems in (\ref{eq:sys}) is said to be \emph{correct} if it satisfies:
\begin{eqnarray} \label{eq:classifier}
f(Y) \,
\left\{
\def\arraystretch{1.3}
\begin{array}{rl}
> 0 & \quad \textrm{if }  Y \in \mathcal L_1 \\
< 0 & \quad \textrm{if }  Y \in \mathcal L_2
\end{array}
\right.
\end{eqnarray}
\qedp
\end{definition}
\salt

The problem of interest is to construct 
correct classifiers. We will investigate two approaches:
\begin{enumerate}
\item[(i)] \emph{Model-based classification}: 
The classifier depends on the knowledge of the matrices
$A_i$ and $C_i$, $i = 1,2$.
\item[(ii)] \emph{Data-driven (model-free) classification}: 
The classifier does not depend on the knowledge of the matrices
$A_i$ and $C_i$, $i = 1,2$, and has to be determined on the basis of
a given number of \emph{sample} trajectories, that is points in the sets $\mathcal L_1$ and $\mathcal L_2$.
\end{enumerate}
Case (i) reflects the situation where the dynamics of the systems are 
known and this information is exploited in the design of the classifier.
On the contrary, case (ii) reflects the situation where the dynamics of the systems are 
unknown or this information is not directly exploited in the design of the classifier.

\subsection{Limitations of the classification problem}

Clearly, the knowledge of the system dynamics provides an extra degree of information
that can be used to properly design a classifier. Yet, there are certain limitations
which cannot be overcome even in the ideal situation where
one has perfect knowledge of the dynamics.
In particular, the following result holds true.
\salt

\begin{theorem}[Limitations of the classification problem] \label{thm:feas}
A correct classifier for the dynamical systems in (\ref{eq:sys}) exists only if 
$\textrm{rank}\, [\, \mathcal O_{1} \,\,\, \mathcal O_{2}\,] = n_{1}+n_{2}$. 
\qedp
\end{theorem}
\salt

\emph{Proof of Theorem \ref{thm:feas}.}
By definition, there exists no correct classifier 
whenever $\mathcal L_1 \cap \mathcal L_2 \ne \emptyset$,
because this implies the existence of trajectories compatible 
with both the systems. This
is equivalent to the fact that the dynamical system 
resulting from the parallel interconnection of $\Sigma_1$ and $\Sigma_2$
is observable, that is that the observability matrix of order $N$
of the pair $(C, A)$ with
\begin{eqnarray} \label{eq:obs_form}
A \,= \, \left[
\begin{array}{cc} \,\, A_{1} \,\, & \,\, 0
\\ \,\, 0 \,\, & \,\, A_{2}
\end{array} \right], \quad 
C \, = \, \left[
\begin{array}{cc} \, C_{1} \, & \, C_{2}
\end{array} \right]
\end{eqnarray}
has column rank $n_1+n_2$. This gives the result. \qedp
\salt

The condition in Theorem \ref{thm:feas} can be satisfied only 
if the two systems do not share common eigenvalue-eigenvector 
pairs. This condition also 
requires $Nm \geq n_1+n_2$, which means
that a large enough observation window must be chosen
to render classification feasible. 

We will take 
this condition as a standing assumption.
\salt

\emph{Assumption 1.} 
$\textrm{rank}\, [\, \mathcal O_1 \,\,\, \mathcal O_2\,] = n_1 + n_2$. \qedp
\salt

\section{Support Vector Machines} \label{sec:stat_learning}

In this section, we briefly recall some concepts on SVM
focusing on the case of \emph{separable} data.
This material of this section is adapted from \cite{Burges1998}.  

Assume we have $L$ observations $(Y_k,\ell_k)$, $k=1,2,\ldots,L$, each one
consisting of a vector $Y_k \in \mathbb R^d$ plus a \emph{label} 
$\ell_k \in \{-1,1\}$ specifying the class to which $Y_k$ belongs.
In connection with the problem introduced in Section II, one can think
of $(Y_k,\ell_k)$ as an observation collected from one of the two
candidate systems $\Sigma_i$, where $Y_k$ is the measurement 
and $\ell_k$ specifies which of the two systems has generated $Y_k$.
Consider the problem of classifying the vectors $Y_k$ using 
hyperplanes $H(Y,\alpha) = \{Y \in \mathbb R^d:\,  w^\top Y + b = 0\}$, where $\alpha=(w,b)$ is 
a vector of adjustable weights.
If there exists a vector $\alpha$ satisfying
{\setlength\arraycolsep{2pt}
\begin{eqnarray} \label{eq:sep_hyp}
\left\{
\def\arraystretch{1.7}
\begin{array}{ll}
w^\top Y_k + b > 0 & \quad \textrm{if } \, \ell_k =1 \\ 
w^\top Y_k + b <  0  & \quad \textrm{if } \, \ell_k = -1 \\
\end{array}
\right. 
\end{eqnarray}}%
for $k=1,2,\ldots,L$, 
then the vectors $Y_k$ are called \emph{linearly separable}, 
and the function 
{\setlength\arraycolsep{2pt}
\begin{eqnarray} \label{eq:lin_classif}
f(Y,\alpha) = w^\top Y + b, \quad \alpha=(w,b)
\end{eqnarray}}%
defines a linear classifier which is correct with respect to the data
$(Y_k,\ell_k)$, $k=1,2,\ldots,L$.
 
For the linearly separable case, an SVM
searches for the separating hyperplane with largest margin $\rho$, that is it
searches for the value of $\alpha$ which maximizes    
{\setlength\arraycolsep{2pt}
\begin{eqnarray} \label{eq:SVM}
\rho := \min_{k=1,2,\ldots,L} \,\, \frac{|w^\top Y_k + b|}{\|w\|}
\end{eqnarray}}%
This can be cast as a convex program:
{\setlength\arraycolsep{2pt}
\begin{eqnarray} \label{eq:SVM_convex}
\def\arraystretch{1.7}
\begin{array}{l}
\displaystyle \min_{\alpha} \, \frac{1}{2} \|w\|^2 \\ 
\textrm{subject to } 
\left\{
\def\arraystretch{1.7}
\begin{array}{ll}
w^\top Y_k + b \geq 1 & \quad \textrm{if } \, \ell_k =1 \\ 
w^\top Y_k + b \leq -1  & \quad \textrm{if } \, \ell_k = -1 \\
\end{array}
\right. 
\end{array}
\end{eqnarray}}%
The reason to search for the hyperplane with largest margin
is related to the fact that $f(Y,\alpha)$ is obtained from a finite 
set of observations, the so-called \emph{training set}. On the other hand,
one would like $f(Y,\alpha)$ to be able to correctly classify also data
which are not present in the training set. This property is usually called  
the \emph{generalization} performance \cite{Burges1998}, and SVM
can guarantee a good generalization performance. We will discuss
this point in more detail in Section V.
\salt

Problem (\ref{eq:SVM_convex}) involves $L$ constraints and $d+1$ unknowns.
When $d > L$ it can be more convenient to resort to a 
dual formulation of the problem, called \emph{Wolfe dual}:
{\setlength\arraycolsep{2pt}
\begin{eqnarray} \label{eq:Wolfe_dual}
\def\arraystretch{1.5}
\begin{array}{l}
\displaystyle \max_{\mu} \,\, \mu^\top \mathbf{1} 
-\frac{1}{2}  \mu^\top Z \mu \\ 
\textrm{subject to } \,\, \mu \succeq 0, \, \displaystyle \sum_{k = 1,2,\ldots,L}  \mu_k  \ell_k = 0
\end{array}
\end{eqnarray}}%
where $\mu := \textrm{col} (\mu_1,\mu_2,\ldots,\mu_{L})$
is the vector of Lagrange multipliers, $Z=[Z_{kj}]$ is a symmetric $L \times L$ 
matrix such that $Z_{kj} = \ell_k \ell_j Y_k^\top Y_j$, $k,j=1,2,\ldots,L$,
and where $\mathbf{1}$ is the vector of ones. 
Problem (\ref{eq:Wolfe_dual}) involves $L$ constraints and unknowns. 
The solution has the form
{\setlength\arraycolsep{2pt}
\begin{eqnarray} \label{eq:supp_vec}
w = \sum_{k = 1,2,\ldots,L} \mu_k \ell_k Y_k
\end{eqnarray}}%
and each vector $Y_k$
associated to a positive multiplier $\mu_k$ is a \emph{support vector}.
This means that for the optimal linear classifier resulting from (\ref{eq:SVM_convex})
the parameter $w$ is given by a linear combination of the support vectors, which can be then 
interpreted as the most representative points in the training dataset.

\subsection{Kernel functions}

Finding a separating surface which is \emph{linear} with respect to the space 
$\mathcal Y$ of the input data is not always possible.
One of the most important results about SVM is related to the 
possibility of finding \emph{non-linear} separating surfaces in a very straightforward manner.

By looking at the optimization problem (\ref{eq:Wolfe_dual}), one sees that  
the data appears only through the products $Y_k^\top Y_j$. One can think of
mapping the input space $\mathcal Y$ into a higher-dimensional space $\mathcal H$
through a function $\Phi: \mathcal Y \rightarrow \mathcal H$, and search for a function 
$\kappa: \mathcal Y \times \mathcal Y \rightarrow \mathbb R$
such that 
\begin{eqnarray} \label{eq:kernel_def}
\kappa (Y, Z) = \langle  \Phi(Y),\Phi(Z)  \rangle, \quad \forall \, X,Z \in \mathcal Y
\end{eqnarray}
The space $\mathcal H$ is called the \emph{feature space}, while
$\Phi(Y)$ is called the \emph{feature vector}.
Any function $\kappa$ satisfying (\ref{eq:kernel_def}) is called a \emph{kernel} function.
Kernel functions define separating 
surfaces which are {linear} with respect to $\mathcal H$.
The remarkable feature of kernel functions is that there is no need 
to use or know the function $\Phi$ in order to compute or use $w$. 
In fact, in order to compute the solution of (\ref{eq:Wolfe_dual}) with respect to $\Phi$
one can simply use $Z_{kj} = \ell_k \ell_j \kappa(Y_k, Y_j)$. Moreover, 
{\setlength\arraycolsep{2pt}
\begin{eqnarray} \label{}
w^\top \Phi(Y) = \sum_{k=1,2,\ldots,L} \mu_k \ell_k  \kappa (Y_k,Y)
\end{eqnarray}}%
Thus one can use $\kappa$ instead of $\Phi$ also for the classification task. 
Kernel functions are also advantageous from the point of view of computations
since $\kappa$ operates in the input space $\mathcal Y$, which has usually lower 
dimension than $\mathcal H$. Common kernel functions are 
\emph{polynomial}, \emph{Gaussian} and \emph{hyperbolic tangent} 
functions \cite{Burges1998}.

In the sequel, we will show that for classifying dynamical systems 
\emph{polynomial} kernels are good candidates. 

\section{Model-based Classification} \label{sec:MB_class}

We now consider a model-based approach to classification.
The following example shows that no correct
classifier exists which is \emph{linear} in the input space.
\salt

\begin{figure}[t] 
\begin{tikzpicture}[scale=1.5]
    \draw [<->,thick] (0,1) node (yaxis) [above] {\small $y(1)$}
        |- (1,0) node (xaxis) [right] {\small $y(0)$};
    \draw [thick] (0,-1) node (yaxism) [below] {}
        |- (-1,0) node (xaxism) [left] {};
    \draw [->,thick] (3,0) -- (4.5,0);
     \draw [thick]  (1.7,0) -- (3,0);
    \draw (-1,-1) coordinate (a_1) -- (1,1) coordinate (a_2);
    \draw (-1,1) coordinate (b_1) -- (1,-1) coordinate (b_2);
    \draw  [thick]  (3.05,-0.05) -- (3.05,0.05);
    \draw  (3.05,-0.1) -- (3.05,-0.1) node [below] {\small $0$};
    \draw  (4.35,-0.1) -- (4.35,-0.1) node [below] {\small $f(Y)$};
    \coordinate (p1) at (0.2,0.2); \coordinate (p2) at (0.4,0.4); \coordinate (p3) at (0.6,0.6);
    \coordinate (p4) at (0.8,0.8); \coordinate (p5) at (-0.2,-0.2); \coordinate (p6) at (-0.4,-0.4);
    \coordinate (p7) at (-0.6,-0.6); \coordinate (p8) at (-0.8,-0.8); 
    \coordinate (n1) at (-0.2,0.2); \coordinate (n2) at (-0.4,0.4); \coordinate (n3) at (-0.6,0.6);
    \coordinate (n4) at (-0.8,0.8); \coordinate (n5) at (0.2,-0.2); \coordinate (n6) at (0.4,-0.4);
    \coordinate (n7) at (0.6,-0.6); \coordinate (n8) at (0.8,-0.8);
     \coordinate (apn) at (2.9,0);
    \coordinate (ap1) at (3.3,0); \coordinate (ap2) at (3.6,0); \coordinate (ap3) at (3.9,0);
    \coordinate (ap4) at (4.2,0);
        \coordinate (an4) at (2.8,0); 
        \coordinate (an3) at (2.5,0); \coordinate (an2) at (2.2,0); \coordinate (an1) at (1.9,0); 
    %
    \fill[red] (p1) circle (1.2pt); \fill[red] (p2) circle (1.2pt); \fill[red] (p3) circle (1.2pt); \fill[red] (p4) circle (1.2pt);
    \fill[red] (p5) circle (1.2pt); \fill[red] (p6) circle (1.2pt); \fill[red] (p7) circle (1.2pt); \fill[red] (p8) circle (1.2pt);
    \fill[blue] (n1) circle (1.2pt); \fill[blue] (n2) circle (1.2pt); \fill[blue] (n3) circle (1.2pt); \fill[blue] (n4) circle (1.2pt);
    \fill[blue] (n5) circle (1.2pt); \fill[blue] (n6) circle (1.2pt); \fill[blue] (n7) circle (1.2pt); \fill[blue] (n8) circle (1.2pt);
    \fill[red] (ap1) circle (1.2pt); \fill[red] (ap2) circle (1.2pt); \fill[red] (ap3) circle (1.2pt); \fill[red] (ap4) circle (1.2pt); 
    \fill[blue] (an1) circle (1.2pt); \fill[blue] (an2) circle (1.2pt); \fill[blue] (an3) circle (1.2pt); \fill[blue] (an4) circle (1.2pt); 
\end{tikzpicture}
\caption{\emph{Left}: Pictorial representation of the possible observation 
points for Example 1 when $N=2$. The trajectories that can be generated 
by the first system correspond to points (red circles) which 
always falls in the first or third quadrant of the 
Cartesian plane, while the trajectories that can be generated 
by the second system correspond to points (blue circles) which 
always falls in the second or fourth quadrant of the Cartesian plane.
\emph{Right}: Pictorial representation of $f(Y):=y(0)y(1)$.}
\label{fig:example}
\end{figure}
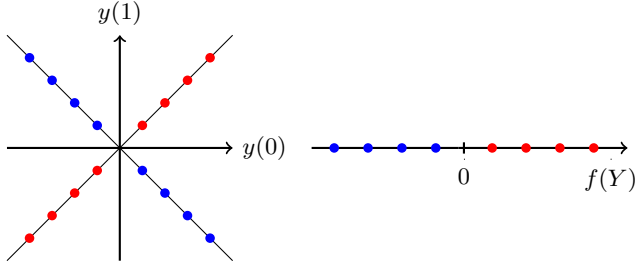

\emph{Example 1.} Consider two systems as in (\ref{eq:sys}), where 
$A_1=1$, $A_2=-1$ and $C_1=C_2=1$. 
Assumption 1 clearly holds true for $N \geq 2$. However, as depicted in Figure \ref{fig:example}, 
there exists no linear classifier for the two candidate systems,
and this is independent of the particular choice of $N$. \qedp
\salt

Example 1 indicates that for dynamical systems there is no correct classifier 
which is linear with respect to $Y$. However, 
Figure \ref{fig:example} suggests that
a correct classifier for Example 1 exists and is given by
$f(Y)= y(0)y(1)$, which can be rewritten as 
$f(Y)=w^\top \Phi$, where 
\begin{eqnarray} \label{eq:feature_example}
w^\top = \frac{1}{2} \left[
\begin{array}{c}
0\\
1 \\
1 \\
0 \\
\end{array}
\right ], \quad \Phi = Y \otimes Y
\end{eqnarray}
where $\otimes$ stands for Kronecker product. We will see that 
$\Phi$ defines a \emph{polynomial} kernel. Before doing this, we show that 
the choice $\Phi = Y \otimes Y$ is general in the sense that it applies 
to any linear dynamical system.

Let $\mathcal G_{i} := \mathcal O^\top_{i}   \mathcal O_{i}$, $i=1,2$,
be the observability Gramian corresponding to the $i$-th system. 
Note that $\mathcal G_{i}$ is nonsingular under Assumption 1.
Also, let 
\begin{eqnarray} \label{eq:Q}
\mathcal Q := \mathcal Q_{1} - \mathcal Q_{2} 
\end{eqnarray}
where
\begin{eqnarray} \label{eq:Q_i}
\mathcal Q_i := \mathcal O_{i} \mathcal G_{i}^{-1} \mathcal O_{i}^\top, \quad i=1,2
\end{eqnarray}
The following result holds true.
\salt

\begin{theorem}[Model-based classifier] \label{thm:MB_classifier}
Let $ \Phi = Y \otimes  Y$ and let $w_M =  \textrm{vec}(\mathcal Q)$, where
$\textrm{vec}(\cdot)$ is the vectorization operator. 
Under Assumption 1, $f(Y) = w_M^\top \Phi$ is a {correct} classifier
for the dynamical systems in (\ref{eq:sys}).
\end{theorem}  
\salt

\emph{Proof of Theorem \ref{thm:MB_classifier}.}
The idea is to show that computing $w_M^\top \Phi$
is equivalent to determining which of the sets $\mathcal L_i$
the vector $Y$ belongs to. Consider the point-set distance
{\setlength\arraycolsep{2pt}
\begin{eqnarray}
\pi_{i}(Y) := \min_{x \in \mathbb R^n} \|\mathcal O_{i} x-Y\|^2, \quad i=1,2
\end{eqnarray}}%
Notice that if $Y \in \mathcal L_1$ then
$\pi_{1}(Y) =0$ and $\pi_{2}(Y) > 0$ in view of Assumption 1. 
Likewise, if $Y \in \mathcal L_2$ then $\pi_{1}(Y) > 0$ and $\pi_{2}(Y) = 0$. 
Hence, the function
\begin{eqnarray} \label{eq:feature_vector_2}
g(Y) := \pi_{2}(Y) - \pi_{1}(Y) 
\end{eqnarray}
defines a {correct} classifier for the dynamical systems in (\ref{eq:sys}).
Notice now that 
$\pi_{i}(Y) = \| (I - \mathcal Q_i^\top) Y \|^2$ for all $Y$. 
Hence, we get
{\setlength\arraycolsep{2pt}
\begin{eqnarray} \label{}
g(Y) &= & Y^\top  (I- \mathcal Q_{2} ) Y - Y^\top (I - \mathcal Q_{1} ) Y  \nonumber \\
&=& Y^\top \mathcal Q Y \nonumber = 
\textrm{vec}(\mathcal Q)^\top ( Y \otimes Y )  \nonumber \\
&=& w_M^\top \Phi
\end{eqnarray}}%
where the first equality follows because $\mathcal Q_i$ is idempotent.
Thus $g(\cdot) = f(\cdot)$, which concludes the proof. \qedp
\salt

\begin{remark}[Invariance to coordinate transformations]
Notice that $w_M$ is independent of the particular state-space 
realization adopted for the dynamical systems since $\mathcal Q_1$
and $\mathcal Q_2$ are projection matrices.
\qedp
\end{remark}

\subsection{Form of the model-based classifier: Kernel function and 
support vectors interpretation}

Theorem \ref{thm:MB_classifier} could have been stated directly in 
terms of $f(Y)=Y^\top \mathcal Q Y$. 
Yet, the form $f(Y)=w^\top \Phi$ turns out to be useful because it provides 
guidelines for the formulation of the data-driven approach.
In fact, it guarantees the existence of a solution for 
the SVM formulation if we use $\Phi = Y \otimes  Y$
as input to the training algorithm. Moreover, it is immediate 
to verify that 
{\setlength\arraycolsep{2pt}
\begin{eqnarray} \label{}
\Phi^\top \Phi = (Y^\top Y)^2
\end{eqnarray}}%
that is $\Phi$ defines a \emph{homogeneous polynomial kernel}.
This means that in the SVM formulation one can
work directly in the space of $Y$ by emplyoing the kernel $\kappa(Y,Z)= (Y^\top Z)^2$.
This option is possible also for the model-based solution
if we write $w_M$ in terms of support vectors. 
This interpretation is simple and worth mentioning.  

A (reduced) singular value decomposition of the matrix $\mathcal O_i$ 
yields $\mathcal O_i = U_i S_i V_i^\top$, where $U_i \in \mathbb R^{Nm \times n_i}$ has orthonormal 
columns, $S_i \in \mathbb R^{n_i \times n_i}$ is a diagonal matrix with positive entries 
(due to Assumption 1), and $V_i \in \mathbb R^{n_i \times n_i}$ is unitary. Thus we have
$\mathcal Q_i = U_i U_i^\top $.
Let $Y_{i,k}$ be the $k$-th column of $U_i$ and let $\Phi_{i,k} = Y_{i,k} \otimes Y_{i,k}$ 
be the corresponding feature vector. Hence,
{\setlength\arraycolsep{2pt}
\begin{eqnarray} \label{}
\sum_{k=1}^{n_i} \Phi_{i,k} &=& \sum_{k=1}^{n_i}  Y_{i,k} \otimes Y_{i,k}  \nonumber \\
&=& \sum_{k=1}^{n_i}  \textrm{vec}( Y_{i,k} Y^\top_{i,k}  )  \nonumber \\
&=& \textrm{vec}(U_i U_i^\top) \nonumber \\
&=& \textrm{vec}(\mathcal Q_i)
\end{eqnarray}}%
where the second equality follows from the vectorization rule
$\textrm{vec}(ABC)=(C^\top \otimes A)\textrm{vec}(B)$ 
for matrices $A$, $B$ and $C$ of appropriate dimension.
Thus,
{\setlength\arraycolsep{2pt}
\begin{eqnarray} \label{}
w_M &=& \textrm{vec}(\mathcal Q) \nonumber \\
&=& \sum_{k=1}^{n_1} \Phi_{1,k} - \sum_{k=1}^{n_2} \Phi_{2,k}
\end{eqnarray}}%
and the support vectors (\emph{cf.} (\ref{eq:supp_vec})) are  
the left singular vectors of the observability matrices $\mathcal O_1$ and $\mathcal O_2$.

\section{Data-driven classification based on \\
Support Vector Machines} \label{sec:DD_class}

The model-based approach suggests an SVM formulation 
for the problem of classifying data generated by dynamical systems.
We will first describe the SVM formulation and we will then 
make some considerations on the \emph{generalization} performance 
of the solution. An interesting result is that the 
generalization performance of the data-driven classifier can be quantified 
as a function of the dynamics of the systems which generate the training dataset. 

\subsection{Data-driven classification based on SVM}
 
Let $\mathcal L_i^{tr} \subset \mathcal L_i$, $i=1,2$,
be a finite nonempty subset of $\mathcal L_i$ consisting of
all the nonzero trajectories recorded from $\Sigma_i$.
Thus $Y_k \in \mathcal L^{tr} := (\mathcal L_1^{tr} \cup \mathcal L_2^{tr})$, 
$k = 1,2,\ldots,L$, $L:=|\mathcal L^{tr}|$,
is the $k$-th training vector. Let $\ell_k=1$ if $Y_k \in \mathcal L_1^{tr}$, and 
$\ell_k=-1$ if $Y_k \in \mathcal L_2^{tr}$.
Finally, let $\Phi_k = Y_k \otimes Y_k$ be
the feature vector associated to $Y_k$. 
Following Section \ref{sec:stat_learning} and Theorem \ref{thm:MB_classifier}, we formulate the 
data-driven approach as the problem 
of finding the hyperplane that contains the origin and 
separates the {training} datasets with maximum margin,
that is:
{\setlength\arraycolsep{2pt}
\begin{eqnarray} \label{eq:SVM}
\def\arraystretch{2.5}
\begin{array}{l}
\displaystyle \min_{w} \, \frac{1}{2} \|w\|^2 \\
\textrm{subject to } \,\,
\left\{
\def\arraystretch{1.7}
\begin{array}{ll}
w^\top \Phi_k \geq 1 & \quad \textrm{if }  \ell_k =1 \\
w^\top \Phi_k \leq -1  & \quad \textrm{if }  \ell_k = -1 \\
\end{array}
\right. 
\end{array} 
\end{eqnarray}}%
The following result holds true. 
\salt

\begin{theorem}[Data-driven classifier] \label{thm:DD_classifier}
Let Assumption 1 be satisfied, and consider an arbitrary training dataset $\mathcal L^{tr}$. Then,   
the solution $w_D$ to the optimization problem (\ref{eq:SVM}) exists and is unique.
Hence, $f(Y) = w_D^\top \Phi$ is a {correct} classifier with respect to $\mathcal L^{tr}$.
\end{theorem}  
\salt 

\emph{Proof of Theorem \ref{thm:DD_classifier}}. 
The proof follows from Theorem \ref{thm:MB_classifier}. In fact,
the model-based solution $w_M$ guarantees that 
$a_1 :=\min_{\ell_k=1} w_M^\top \Phi_k > 0$
and $a_2 :=\max_{\ell_k=-1} w_M^\top \Phi_k < 0$.
Thus $\overline w_M := w_M/a$ with
$a :=\min \{a_1,-a_2\}$ guarantees the
feasibility of the set of constraints.
Uniqueness follows as the optimization problem
is a convex program. 
\qedp
\salt

The constraint that the solution must contain the origin is 
simply to mimic the model-based solution. This constraint 
is actually not needed, and a standard formulation (\ref{eq:SVM_convex}) 
would still guarantee existence and uniqueness of the solution.
If we constrain the solution to contain the origin the Wolfe dual 
becomes
{\setlength\arraycolsep{2pt}
\begin{eqnarray} \label{eq:Wolfe_dual_simple}
\def\arraystretch{1.5}
\begin{array}{l}
\displaystyle \max_{\mu} \,\, \mu^\top \mathbf{1} 
-\frac{1}{2}  \mu^\top Z \mu \\
\textrm{subject to } \,\, \mu \succeq 0 
\end{array}
\end{eqnarray}}%
which does not
involve the constraint $\sum_{k=1,2,\ldots,L} \mu_k  \ell_k = 0$.
We notice that in this case the dimension of the feature space is $(Nm)^2$.
Nonetheless, by considering a kernel-based implementation one can remain in the 
$Nm$-dimensional space of the sequences $Y$. 

Theorem \ref{thm:DD_classifier} indicates that one can find 
a surface separating the training dataset without information about
the underlying systems except for their linearity, which suggests 
the feature space of choice. In the remaining part of this section,
we will discuss on the capability of this SVM classifier to \emph{generalize} 
to observations outside the training set. 

\subsection{Expected risk}

Ideally, one would like to establish the correctness of the data-driven
classifier in the same sense as Definition \ref{def:class}. This is a non-trivial 
problem which, to the best of our knowledge, has not yet been solved.  
In the sequel, we consider another way to characterize the 
{generalization performance} of the SVM
classifier, based on the notion of \emph{expected risk}. While this 
notion does not provide deterministic bounds, it has the merit to capture 
the situation where the training dataset is randomly chosen. 
Hence, it has the merit to describe cases in which one cannot perform dedicated 
experiments on the systems. 

Consider a training dataset of $L$ \emph{random i.i.d.} observations drawn according to
a probability distribution $P(Y,\ell)$. Given a classifier $f(Y)$, its \emph{expected risk}
can be defined as \cite{Vapnik1995}:
{\setlength\arraycolsep{2pt}
\begin{eqnarray}
R :=  \int \frac{1}{2} \left| \ell - \textrm{sgn} ( f(Y) ) \right| dP(Y,\ell)
\end{eqnarray}}%
where $\textrm{sgn}$ is the sign function. The expected risk quantifies 
the capability of a classifier to generalize from the training dataset. 
Several studies have been devoted to provide upper bounds
on the expected risk for a given family of classifiers. 
An interesting  bound for linear classifiers which serves our discussion is reported
hereafter.
\salt

\begin{theorem}[\cite{Bartlett1999}] \label{thm:GTC_v}
Let $Y \in \mathbb R^d$ belong to the sphere of 
radius $R$, and consider the class $\mathcal F$ of real-valued functions defined as 
$\mathcal F := \{Y \mapsto w^\top Y: \|w\| \leq 1, \|Y\| \leq R\}$. There is a constant $c$
such that, for all probability distributions, with probability at least $1- \eta$ over $L$
randomly i.d.d. vectors, if a classifier has margin at least $\rho$ on all the examples
then its expected error is not larger than
{\setlength\arraycolsep{2pt}
\begin{eqnarray} \label{eq:GTC_h_DD}
\frac{c}{L} \left( \frac{R^2}{\rho^2} \log^2 L + \log \left( 1/\eta \right) \right)
\end{eqnarray}}%
\qedp
\end{theorem}
\salt  

The constant $c$ is related to the so-called \emph{fat-shattering}
dimension of linear classifiers, and its explicit expression can be
found in \cite{Bartlett1999}.
Theorem \ref{thm:GTC_v} shows that one can quantify the generalization
performance of a linear classifier as a function of the margin $\rho$ obtained 
for the training dataset. We now show that the margin of the SVM 
classifier can be related to the margin of the model-based solution.
This permits to quantify the generalization performance of the SVM classifier
in terms of the dynamics of the systems that one wishes to classify. 
The analysis which follows holds for normalized data. We will briefly comment
later on the general case.

Consider normalized training data
{\setlength\arraycolsep{2pt}
\begin{eqnarray} \label{eq:norm_data}
\overline Y_k := \frac{Y_k}{\|Y_k\|} 
\end{eqnarray}}%
with feature vector $\overline \Phi_k := \overline Y_k \otimes \overline Y_k$. It holds that
$\| \overline \Phi_k \| =1$. Consider now the optimal solution $w_D$ 
to (\ref{eq:SVM}) computed with respect to $\overline \Phi_k$,
whose existence and uniqueness is again ensured by the model-based solution.
We can assume without loss of generality that $\|w_D\| \leq 1$.
Let now $\rho_M$ and $\rho_D$ represent the margin corresponding 
to the model-based and the data-driven solutions, 
respectively, 
{\setlength\arraycolsep{2pt}
\begin{eqnarray} \label{eq:MB_DD_margin}
\rho_i := \min_{k=1,2,\ldots,L} \, \frac{ |w_i^\top \overline \Phi_k |}{\|w_i\|}, \quad i \in \{M,D\}
\end{eqnarray}}%
It holds that
{\setlength\arraycolsep{2pt}
\begin{eqnarray} \label{eq:rel_MB_DD}
\rho_D \geq \rho_M
\end{eqnarray}}%
irrespective of the training dataset, 
because the data-driven solution is the margin maximizer. 
The next result shows that, using normalized data, $\rho_M$ is bounded from below by a positive 
quantity that depends solely on the dynamics of the systems one wishes to classify. 
We refer the reader to \cite{Cock2002} for a definition of 
\emph{principal angles}.
\salt 

\begin{theorem}[Bound on the data-driven classifier margin] \label{thm:DD_margin}
Let Assumption 1 be satisfied. Consider an arbitrary training dataset 
of vectors $Y_k$, and let $\overline \Phi_k := \overline Y_k \otimes \overline Y_k$ 
where $\overline Y_k$ is as in (\ref{eq:norm_data}). Let
$w_D$ be the unique solution to the optimization problem (\ref{eq:SVM})
computed with respect to $\overline \Phi_k$.
Then, it holds that
{\setlength\arraycolsep{2pt}
\begin{eqnarray}
\rho_D \geq \frac{\beta}{\sqrt{2 \,(n_1+n_2)}}
\end{eqnarray}}%
where $n_1$ and $n_2$ are the 
orders of the dynamical systems in (\ref{eq:sys}), and
$\beta$ is 
the squared sine of the smallest principal angle between 
the subspaces spanned by the columns of the 
observability matrices $\mathcal O_1$ and $\mathcal O_2$.
\end{theorem}  
\salt
 
\emph{Proof of Theorem \ref{thm:DD_margin}.}
Since $\rho_D \geq \rho_M$ it is sufficient to bound $\rho_M$. 
The term $\|w_M\|$ satisfies
{\setlength\arraycolsep{2pt}
\begin{eqnarray} \label{}
\|w_M\|^2 &=& \textrm{vec}(\mathcal Q)^\top \textrm{vec}(\mathcal Q) \nonumber \\
 &=& \|\mathcal Q\|_F^2 \nonumber \\
 &\leq& 2 \,(n_1+n_2)
\end{eqnarray}}%
where $\|\cdot\|_F$ denotes Frobenius norm. The 
inequality follows from 
$\|\mathcal Q\|_F^2 \leq 2 \|\mathcal Q_1\|_F^2 + 2 \|\mathcal Q_2 \|_F^2$
and $\|\mathcal Q_i\|_F^2=n_i$ because the $\mathcal Q_i$'s are
projection matrices. 
Consider now the term $|w_M^\top \overline \Phi_k|$. Assume without loss of 
generality that its minimum is attained for some $Y_* \in \mathcal L_1^{tr}$.
It holds that 
{\setlength\arraycolsep{2pt}
\begin{eqnarray} \label{}
\min_{k=1,2,\ldots,L} | w_M^\top \overline \Phi_k | &=& |w_M^\top \overline \Phi_* | \nonumber \\
&=& | \pi_2(\overline Y_*) - \pi_1(\overline Y_*) | \nonumber \\
&=& \pi_2(\overline Y_*) 
\end{eqnarray}}%
The third equality comes from the fact that $Y_* \in \mathcal L_1^{tr}$ implies 
$\pi_2(\overline Y_*)>0$ and $\pi_1(\overline Y_*)=0$ in view of Assumption 1.
As shown in \cite[Theorem 1]{battSCL}, 
$\pi_2(\overline Y_*) \geq {\beta} \|\overline Y_*\|^2$.
Hence, the proof follows from $\|\overline Y_*\| =1$.
\qedp
\salt

Theorem \ref{thm:DD_margin} permits to bound the risk of the data-driven classifier based on
the dynamics of the systems one wishes to classify, and formalizes the 
intuition that the risk bound becomes smaller as the dynamics of the 
systems to classify are more distant from one another. In fact, 
the higher $\beta$ the larger the coefficient of inclination between the 
subspaces spanned by the columns of $\mathcal O_1$ and $\mathcal O_2$, 
which is maximal when the two spaces are orthogonal. 

Data normalization ensures that $\rho_M$ is bounded away from zero.
This property does not hold in general since trajectories of dynamical systems can be 
arbitrarily close to the origin. Nonetheless, one can 
obtain a very similar bound by adding to (\ref{eq:GTC_h_DD}) an
extra term which accounts for training data below the margin $\rho$
\cite[Theorem 1.7]{Bartlett1999}.

\section{Discussion}

It is intuitive that incorporating models can be
beneficial to the classification task. This fact is obvious also from the analysis 
shown in this paper since under Assumption 1 no classifier can outperform 
the model-based classifier when the models are exact and 
the data are noise-free. However, as mentioned before,
the model-based approach introduces the non-trivial issue of how to quantify the effect of 
modelling inaccuracies. The data-driven bypasses the intermediate step of identification, 
and thus it has the potential to be applicable also when accurate models
are difficult to obtain. Hereafter, we briefly elaborate on this point also in connection 
with a number of open problems.

\subsection{Linear systems with noisy observations}

When observations are corrupted by noise, identification 
may be difficult and require, even for linear systems, many careful provisions \cite{Pillonetto14}.
In contrast, an SVM formulation can address the problem in a rather straightforward manner. 
Consider a \emph{soft-margin} SVM \cite{Burges1998}:
{\setlength\arraycolsep{2pt}
\begin{eqnarray} \label{eq:SVM_soft}
\def\arraystretch{2.5}
\begin{array}{l}
\displaystyle \min_{(w,\xi)} \, \frac{1}{2} \|w\|^2 + \sum_{k=1,2,\ldots,L} C\, \xi_k \\
\textrm{subject to } \,\,
\left\{
\def\arraystretch{1.7}
\begin{array}{ll}
w^\top \Phi_k \geq 1 - \xi_k & \quad \textrm{if }  \ell_k =1 \\
w^\top \Phi_k \leq -1 + \xi_k  & \quad \textrm{if }  \ell_k = -1\\
\end{array}
\right. 
\end{array} 
\end{eqnarray}}%
where $C$ is a parameter and $\xi := \textrm{col }(\xi_1,\xi_2,\ldots,\xi_L)$ is the vector 
of \emph{slack} variables, which account for the fact that noise may render the 
data non-separable. On one hand, there exist many studies  
aimed at quantifying the generalization performance of SVM also for 
soft-margin formulations \cite{Taylor2002}. On the other hand, even with noisy data
one can still give a separation measure between linear systems (the margin $\rho_M$)  
as a function of their dynamics and 
the \emph{signal-to-noise ratio} \cite[Theorem 2]{battSCL}. This means 
that even with noisy data
one can quantify the generalization performance of an SVM classifier along the same lines 
as in Section V-B.

We point out that while this reinforces the idea
that for linear systems \emph{polynomial} kernels
are good candidates, it remains unclear if better performance 
can be obtained with different kernel functions. 


\subsection{Nonlinear systems}

Classification for nonlinear systems is another situation in which an SVM formulation 
can bypass difficulties related to system
identification. This is related to the capability of SVM to find 
non-linear separating surfaces in a straightforward manner through the kernel trick.
Interestingly, even in the nonlinear case
one can define a separation measure between dynamics \cite[Theorem 3]{battTAC}. 
However, in contrast with the linear case where this measure
involves principal angles between observability subspaces,
for nonlinear systems this measure involves \emph{$\mathcal K$-functions},
which are often difficult to relate to the underlying dynamics.
Like for linear systems, a deeper understanding of this point would be beneficial to figure out 
which types of kernel functions are most suitable for a given class of nonlinear systems.

In fact, theoretical studies on classification for nonlinear systems are 
recent also within computer science, and the approaches appear largely diversified; for example,
see \cite{Shen2017} for an interesting recent account. Yet, also in this context,
the question of which kernel functions are most suitable for a given class of dynamics
is unresolved. 

\subsection{Classifiers in-the-loop}

Thanks to their simple form, classifiers have the potential to be used in 
real-time applications, thus for control purposes.
This fact has been noted in \cite{Poonawala2017}, where the authors 
introduce the term \emph{classifier in-the-loop} to describe a framework 
in which a classifier can modify online the control action by looking at the process data. 
A notion of generalization is considered, which characterizes the capability of a classifier to work under 
small perturbations of the system vector fields, which is a \emph{sensitivity}-type 
analysis. While the results are promising, it remains unexplored how to handle more general forms 
of uncertainty. Ideally, one should provide bounds on the risk 
function of a classifier that hold for all the possible system trajectories, 
and relate such bounds with closed-loop stability properties.
A non-trivial difficulty is that much of the theory on the generalization 
properties of classifiers have been developed in a probabilistic setting,
while for robust stability it is desirable to guarantee worst-case deterministic bounds.

Interestingly, the architecture considered in \cite{Poonawala2017}
can be regarded as a \emph{supervisory} control system \cite{liberzon}.
In supervisory control, the supervisor selects based on process data
which candidate control law (\emph{hypothesis}) is most appropriate at any given time. This is done by assigning to each candidate 
law a score function (\emph{cost function}) that quantifies the performance level 
achievable by the control law given the process data,
In supervisory control, one often uses the term \emph{cost detectability} \cite{safo08,TAC13} 
to measure the capability of a supervisor to learn from data an appropriate control law 
even when the process does not match the models used to design the control laws.
In fact, the supervisor is a classifier and \emph{cost detectability} is a measure of 
its generalization performance. 
The idea of adaptive control as \emph{learning-from-data}
is indeed not new \cite{safo}, but a firm theoretical link with the realm of machine learning 
has not yet been established. 

\section{A Numerical Example}

Consider a system with transfer function
{\setlength\arraycolsep{2pt}
\begin{eqnarray} \label{}
G(s) = \frac{s+1}{(s+10)(s^2+s+1)}
\end{eqnarray}}%
where $s$ is the Laplace variable. 
To improve performance, the system is controlled with 
a proportional controller $K=30$ under negative feedback.
The goal is to design a classifier which can detect the \emph{loss
of control effectiveness}. We denote by $\overline \Sigma_1$
the open-loop system and by $\overline \Sigma_2$
the closed-loop system. Hence, $\overline \Sigma_1$ and $\overline \Sigma_2$ 
have transfer functions $G(s)$ and $W(s):=KG(s)/(1+KG(s))$, respectively.
Finally, we denote by $\Sigma_1$ and $\Sigma_2$ the corresponding sampled-data systems 
under sampling time $T_s$. The systems are as in (\ref{eq:sys}) with $n_1=n_2=3$
and $m=1$. 
Using the previous notation, we let $N$ be  
the length of the observation sequences, and $L$ the number of training data. 
We let $Q$ be the number of data used for validation. 
In order for Assumption 1 to be satisfied one needs $N\geq6$.
Under such condition, Assumption 1 holds  for a generic choice of $T_s$.

We focus on the SVM classifier because the model-based classifier is always correct 
under Assumption 1. We note that classifying the two systems is non-trivial,
as one can observe from Figure 2. For instance, for $T_s=0.1$, in the ideal case of $N=\infty$
one has $\beta=0.004$ in Theorem \ref{thm:DD_margin},
and a \emph{cepstral} distance \cite{Cock2002} equal to $1.045$.

We report in Table I simulations results for various choices of $N$, $L$ and $T_s$,
with $Q=1000$ validation data. The SVM classifier is computed as in Theorem \ref{thm:DD_margin}. 
For the training and the validation test, trajectories are generated from random initial conditions with zero mean 
and variance $\sigma^2=100$. The error in the validation test is defined as:
{\setlength\arraycolsep{2pt}
\begin{eqnarray}
R_{test} := \sum_{k =1,2,\ldots,Q} \,\,
\frac{1}{2 Q}  \left| \ell_k - \textrm{sgn} ( w_D^\top \overline \Phi_k ) \right| 
\end{eqnarray}}%
\begin{table}[h!]
\centering
\begin{tabular}
{@{}l*5{>{}l}%
l<{Example text}l@{}}
\toprule[1.5pt]
& \multicolumn{5}{l}{\head{Variation of the parameter $N$\, ($T_s=0.1$, $L=50$)}}\\
& \normal{\head{$N=2$}} & \normal{\head{$N=5$}} & \normal{\head{$N=10$}} & \normal{\head{$N=50$}} & \head{$N=100$}\\
\cmidrule(){2-6}
\multirow{1}{*}{$R_{test}$} 
& \normal{$0.4720$} & \normal{$0.0760$} & \normal{$0.0685$} & \normal{$0.0150$} & \normal{$0.0150$}  \\
& \normal{ } & \normal{} & \normal{} & \normal{}  \\
\toprule[1.5pt]
& \multicolumn{5}{l}{\head{Variation of the parameter $L$\, ($T_s=0.1$, $N=10$)}}\\
& \normal{\head{$L=3$}} & \normal{\head{$L=5$}} & \normal{\head{$L=10$}} & \normal{\head{$L=50$}} & \normal{\head{$L=100$}} \\
\cmidrule(lllll){2-6}
\multirow{1}{*}{$R_{test}$} 
& \normal{$0.1480$} & \normal{$0.1480$} & \normal{0.0720} & \normal{$0.0685$} & \normal{0.0620} \\
& \normal{ } & \normal{} & \normal{} & \normal{}  \\
\toprule[1.5pt]
& \multicolumn{5}{l}{\head{Variation of the parameter $T_s$\, ($N=10$, $L=50$)}}\\
& \normal{\head{$T_s=0.01$}} & \normal{\head{$T_s=0.05$}} & \normal{\head{$T_s=0.1$}} & \head{$T_s=0.5$} & \head{$T_s=1$} \\
\cmidrule(lllll){2-6}
\multirow{1}{*}{$R_{test}$} 
& \normal{$0.4850$} & \normal{$0.0730$} & \normal{$0.0685$} & \normal{$0.0385$} & \normal{$0.0835$}  \\
& \normal{ } & \normal{} & \normal{} & \normal{}  \\
\toprule[1.5pt]
\end{tabular}
\caption{Numerical results for the SVM classifier with $Q=1000$.}
\end{table}

\begin{figure*}[h!] 
\begin{center}
\psfrag{samples}{\footnotesize $y(1)$}
\begin{tabular}{ll}
\includegraphics [width=0.46\textwidth] {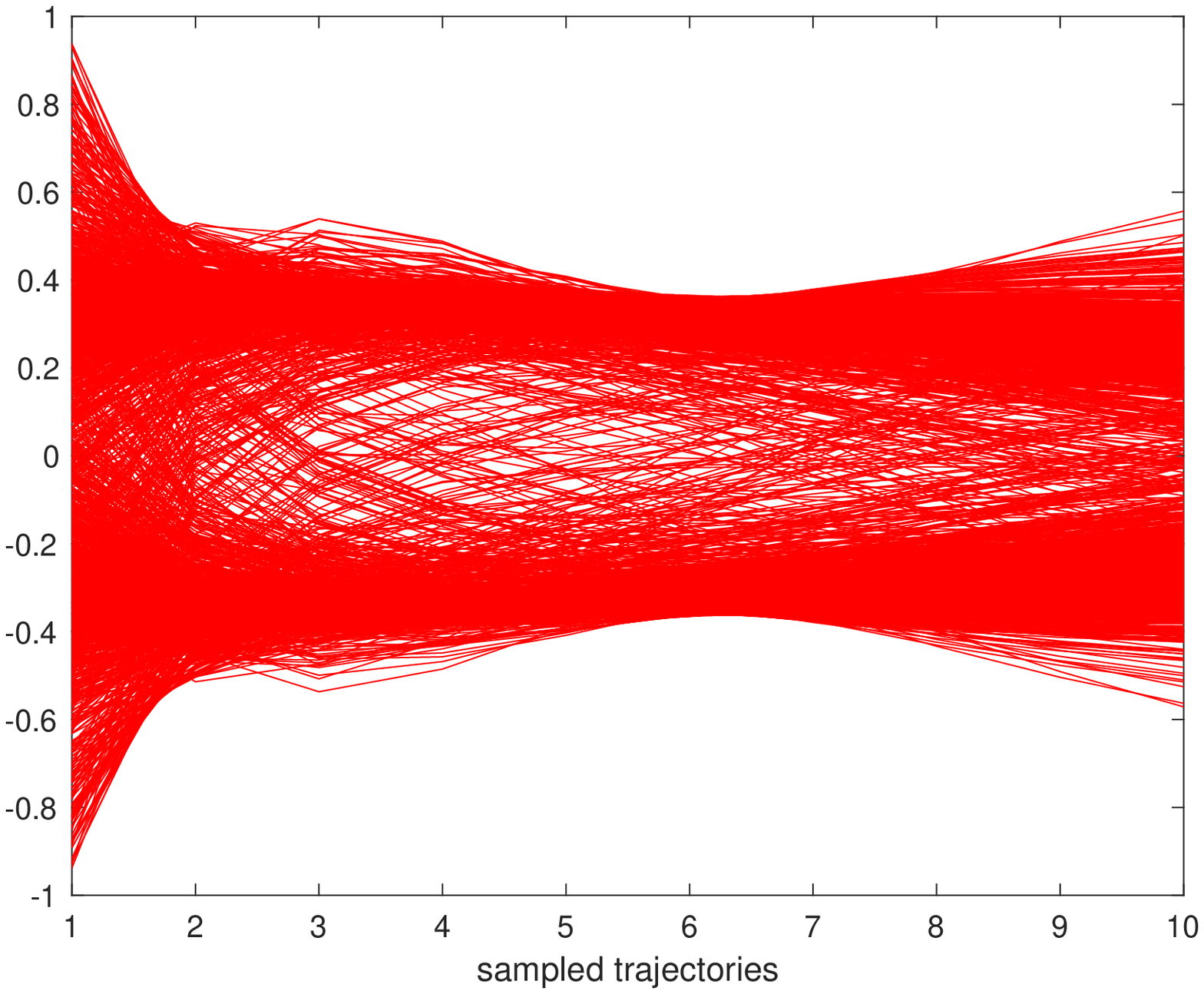} &
\includegraphics [width=0.46\textwidth] {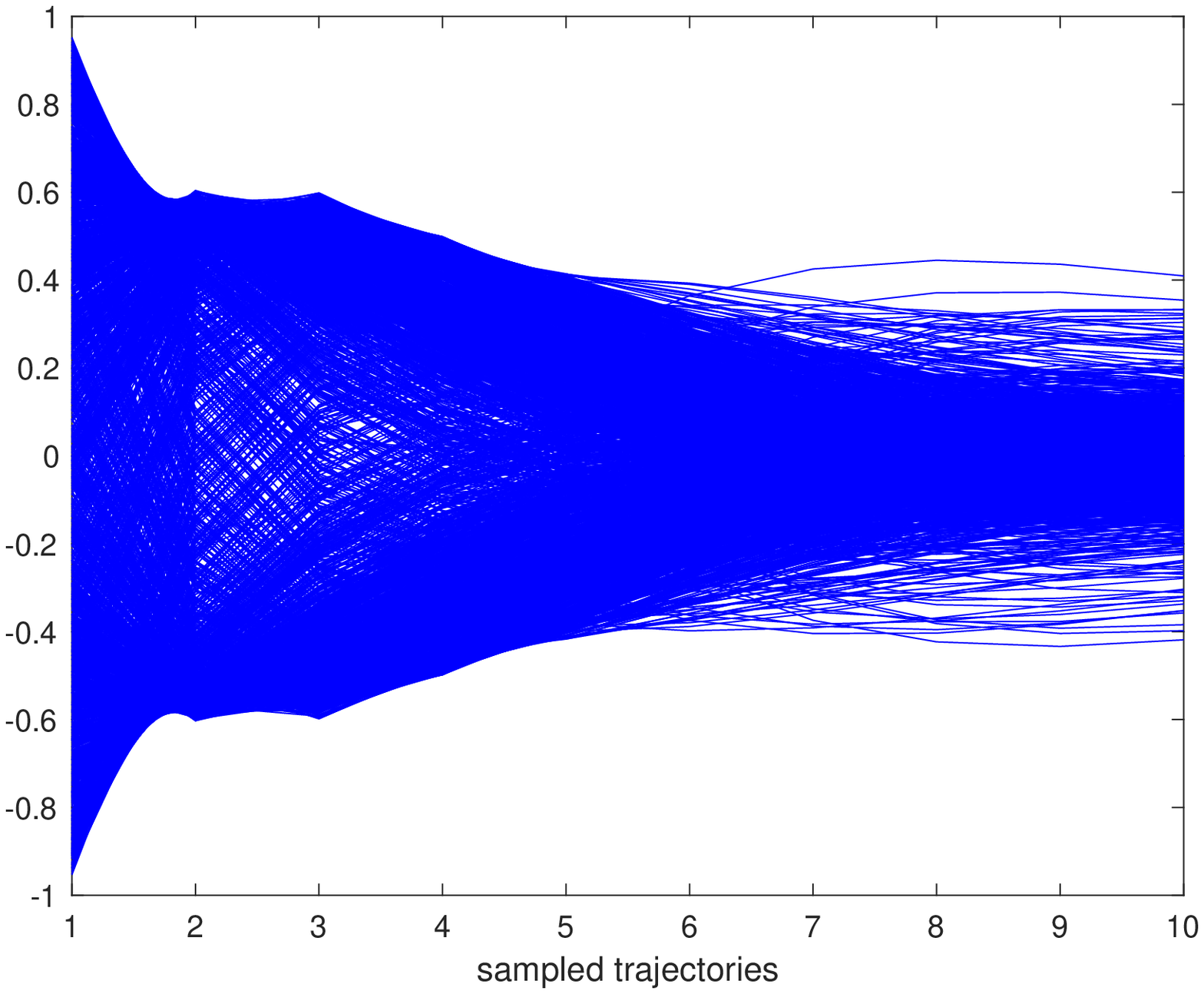}
\end{tabular}
\caption{Output trajectories of the two systems (Left: open-loop; Right: closed-loop) with
$N=10$ and $T_s=0.1$. The figures report $1000$ normalized trajectories for each system 
generated from random initial conditions with zero mean and variance $\sigma^2=100$.}
\end{center} 
\label{fig:example2}  
\end{figure*}

One sees that the classifier performs well for reasonable 
choices of the parameters. In particular:
\begin{enumerate}
\item[(i)] \emph{Dependence on $N$}. As $N$ goes to zero, the performance is clearly that of
a random guess. One the other hand,
remarkably, classification becomes accurate exactly as soon as one approaches the theoretical bound $N\geq6$.
The performance saturates after $N=50$. To further decrease the error we need to increase $L$
(with $L>300$ one can achieve an error below $1\%$).
\item[(ii)] \emph{Dependence on $L$}. The performance variations are less evident in this case.
This suggest that $L$ is less critical than $N$. The intuition is that random initial conditions generically ensure the
excitation of all system dynamics so that even few examples may suffice. 
\item[(iii)] \emph{Dependence on $T_s$}. The sampling time does not play a major role as long as 
we avoid \emph{over-sampling}, in which case both $A_1$ and $A_2$ tend to the identity matrix, 
or \emph{under-sampling}, in which case both $A_1$ and $A_2$ tend to the zero matrix.
\end{enumerate}
 
\section{Concluding Remarks}

We have considered the problem of classifying trajectories generated by dynamical 
systems, looking at a model-based approach, the 
common approach in control engineering, as well as at a data-driven approach based 
on Support Vector Machines, a popular method in computer science.
The present discussion suggests that both the approaches have distinct merits.
A deeper understanding of the interplay between these two approaches
would help to establish a sound theory for dynamical systems 
more general than those considered in this paper.

\bibliographystyle{IEEEtran}

\bibliography{SVMbiblio}   
      
\end{document}